\documentclass[prd,twocolumn,
preprintnumbers,
nofootinbib,
noshowpacs
]{revtex4} %floatfix
\usepackage{epsfig}
\usepackage{color}
\usepackage{amsmath}
\usepackage{amssymb}
\usepackage{natbib}
\usepackage{enumerate}
\usepackage{hyperref}
\usepackage[normalem]{ ulem }
\usepackage{soul}

\usepackage{graphicx}
\usepackage{units}
\graphicspath{{figs/}}

\usepackage[caption=false]{subfig}
\usepackage{slashed}

\usepackage[dvipsnames]{xcolor}

\newcommand{\be}{\begin{equation}}
\newcommand{\ee}{\end{equation}}
\newcommand{\beq}{\begin{equation}}
\newcommand{\eeq}{\end{equation}}
\newcommand{\bea}{\begin{eqnarray}}
\newcommand{\eea}{\end{eqnarray}}
\newcommand{\beaa}{\begin{eqnarray*}}
\newcommand{\eeaa}{\end{eqnarray*}}
\newcommand{\ba}{\begin{array}}
\newcommand{\ea}{\end{array}}
\newcommand{\bi}{\begin{itemize}}
\newcommand{\ei}{\end{itemize}}
\newcommand{\ben}{\begin{enumerate}}
\newcommand{\een}{\end{enumerate}}

\begin{document}
%%%%%%%%%%%%%%%%%%%%%%%%%%%%%%%%%%%%%%%%%%%%%%%%%%%%%%%%%%%%%%%%%%%%%%
\preprint{ULB-TH/16-22}
%\title{New leptogenesis mechanism in the standard seesaw model at low scale}
\title{A Clockwork WIMP}

\author{Thomas Hambye}
\email{thambye@ulb.ac.be}
\author{Daniele Teresi}
\email{daniele.teresi@ulb.ac.be}
\author{Michel H.G. Tytgat}
\email{mtytgat@ulb.ac.be}

\affiliation{Service de Physique Th\'eorique - Universit\'e Libre de Bruxelles, Boulevard du Triomphe, CP225, 1050 Brussels, Belgium}

%\date{\today}

%%%%%%%%%%%%%%%%%%%%%%%%%%%%%%%%%%%%%%%%%%%%%%%%%%%%%%%%%%%%%%%%%%%%%%
\begin{abstract}
We embed a thermal dark matter (DM) candidate within the clockwork framework. 
This mechanism allows to stabilize the DM particle over cosmological time because it suppresses its decay into Standard Model (SM) particles. 
At the same time, pair annihilations are unsuppressed, so that  the relic density is set by the usual freeze-out of the DM particle from the thermal bath. The slow decay of the DM candidate is induced by ``clockwork'' particles that can be quite light (rather than at some GUT or Planck scale) and could be searched for at current or future colliders. According to the scenario considered, the very same particles also mediate the annihilation process, thus providing a connection between DM annihilation and DM decay, and fixing the mass scale of the clockwork states, otherwise unconstrained, to be in the TeV range or lighter. We then show how this setup can minimally emerge from the deconstruction of an extra dimension in flat spacetime.
Finally, we argue that the clockwork mechanism that we consider  could induce Majorana neutrino masses, with a seesaw scale of order TeV or less and Yukawa couplings of order unity.
\end{abstract}

\maketitle
%%%%%%%%%%%%%%%%%%%%%%%%%%%%%%%%%%%%%%%%%%%%%%%%%%%%%%%%%%%%%%%%%%%%%%

\section{Introduction}
A viable dark matter (DM) particle must be stable over cosmological time scales. This requirement generally dictates many properties of a DM candidate. Depending on the model, it could be absolutely stable or very long lived (see {\em e.g.}~\cite{Hambye:2010zb} for a general discussion of the stability issue).  Absolute stability may be due, for instance, to a  gauge symmetry, as for the electron in the SM. Alternatively, DM could be long-lived for accidental reasons, as the proton, whose stability is associated to a global continuous symmetry, such as the baryon number. The possibility that DM may be long lived is actually very interesting as its decay could be probed through indirect-detection searches. Constraints from fluxes of cosmic rays impose that the lifetime of DM candidates in the WIMP mass range, which we will consider in the sequel, must be typically larger than $\tau \sim 10^{26}$~sec. The most straightforward way to get such a large lifetime is to assume that the decay is induced by the exchange of very heavy particles, encapsulated at low energies by  mass-suppressed effective operators. Assuming couplings of order one, instability associated to a dimension-5 interaction would require that the heavy degrees of freedom lie way  above the Planck scale. In contrast, a dimension-6 operator only requires that these particles are around the GUT scale. This is appealing but also means that the nature of these very heavy degrees of freedom would be impossible to test directly. Alternatively, one could assume that the particles that trigger the decay of DM are much lighter, but then with much smaller couplings (something that we are usually reluctant to assume, unless protected from large quantum corrections by a symmetry). However this cannot be tested either. 

In the following we propose that, along a clockwork mechanism \cite{Choi:2015fiu,Kaplan:2015fuy,Giudice:2016yja,Kehagias:2016kzt,Farina:2016tgd,Ahmed:2016viu}, a DM particle could be made accidentally stable but have a decay into SM particles induced by order unity interactions with particles that could be produced at colliders. Such a clockwork mechanism does not prevent the DM to undergo fast annihilations into SM particles (or hidden-sector particles) so that its relic density is determined by the standard freeze-out mechanism, {\em i.e.} it is a WIMP. 

As we will see, we need to introduce a lot of new fields 
%and symmetries 
(and specific couplings). One may right away wonder whether this is worth explaining the mere stability of dark matter. 
%Here we embrace the philosophy of the clockwork mechanism as exposed in \cite{Giudice:2016yja} {\em i.e.} 
However, beside that the potential benefit is to have a whole world of new particles at a rather low scale, implying a rich phenomenology at current and future experiments,
%(with the same clockwork setup responsible for the generation of the other small numbers/hierarchies needed in the different sectors of the theory) 
the presence of all these fields can be the 4-dimensional consequence of theories with extra-dimensions. We will show in particular that a fermion clockwork chain can arise in a simple way from the deconstruction of a single field in flat spacetime.

The plan of the article is as follows. In the next section \ref{sec:clock} we build the model, based on the clockwork mechanism and expose the basic properties of the DM candidate. The basic phenomenology is developed in section~\ref{sec:pheno}. In section~\ref{sec:deconstr} we discuss  how this kind of setup can be minimally obtained from the deconstruction of an extra dimension, whereas in section~\ref{sec:neutrino} we comment on the possibility of inducing Majorana neutrino masses from a similar mechanism. We finally draw our conclusions in section \ref{sec:conc}. 

\section{Clockwork dark matter}
\label{sec:clock}
 One may --- but does not have to --- think of the clockwork mechanism as  arising from the deconstruction \cite{ArkaniHamed:2001ca}  of an extra spatial dimension. The resulting ``theory space'' consists of a series of fields with adjacent interactions. 
  Here we start with  $N$ Dirac fermions $\psi_i \equiv (L_i, R_i)$ (with $i= 1 \ldots N$), similarly to~\cite{Giudice:2016yja}, together with a single right-handed chiral fermion ${R_0}$. We thus have the following set of $(2N+ 1)$ chiral  fields
  \begin{equation}
  {R_0},  {L_1},{R_1},\ldots,  {L_N}, {R_N}\,.
  \end{equation}
  We refer to this set as the clockwork chain. 
  Each chiral fermion is charged under a $U(1)$ chiral symmetry. The total symmetry group is thus
\begin{equation}\label{eq:group}
U(1)_{R_0} \times U(1)_{L_1} \times U(1)_{R_1} \times \ldots\times U(1)_{L_{N}} \times U(1)_{R_{N}} \,,
\end{equation}
and we identify the last abelian factor with the lepton-number symmetry of the SM:
\begin{equation}
U(1)_{R_{N}} \equiv U(1)_{L_{SM}}\,.
\end{equation}
The latter choice makes natural a Yukawa coupling $y$ of ${R_N}$ with the SM left-handed leptons, here denoted by $L_{SM}$. 
We also introduce, as in Ref.~\cite{Giudice:2016yja}, two sets of $N$ spurions $S_i$  and $C_i$ with $S_i \sim (1,-1)$ under $U(1)_{L_{i+1}} \times U(1)_{R_i}$ and $C_i \sim (1,-1)$ under $U(1)_{L_i} \times U(1)_{R_{i}}$, which break the symmetry (\ref{eq:group}) to $U(1)$. 
%They will play the role of the spurions considered in \cite{Giudice:2016yja}. 
With these ingredients, the Lagrangian takes on the form 
\begin{align}
\label{eq:lag1}
\mathcal L &= \mathcal{L}_{\rm SM} + \mathcal{L}_{\rm kinetic} \notag\\
& - \sum_{i=1}^{N} \big(y_S S_i \bar{L}_i R_{i-1} - y_C C_i \bar{L}_i R_{i} + h.c. \big) \notag\\
%&- \sum_{i=1}^{N} \big(\lambda_S \, S_i^\dag S_i \, H^\dag H + \lambda_C \,C_i^\dag C_i \, H^\dag H \big) \notag\\
%&- \sum_{i,j=1}^{N} \lambda_{SC} \, S_i^\dag S_i \,C_j^\dag C_j  \notag\\
& - ( y\,  \bar L_{SM} \widetilde{H} {R}_N + h.c.) -  \frac{1}{2} (m_N \overline{R^c_0}\,R_0 + h.c.)\,.
\end{align}
For simplicity we take all spurions to be universal
\begin{equation}
y_S S_i  = m \;, \qquad y_C  C_i   = M \equiv q\,m  \quad {\rm with}\quad  q > 1\,.
\end{equation}
We have also introduced a soft breaking of $U(1)_{R_{0}}$ which gives a Majorana mass $m_N$ to the light eigenstate $N \sim R_0$ (thus breaking the residual $U(1)$ symmetry). 
The fermionic mass Lagrangian thus reads
\begin{equation}\label{eq:clock_lagr}
\mathcal{L} \supset - \,m\, \sum_{i=1}^{N} \big(\bar{L}_i R_{i-1} - q\, \bar{L}_i R_{i}  \big)  - \frac{m_N}{2}\, \overline{R^c_0}\,R_0 + h.c.
\end{equation}
For the phenomenological analysis in the next section we will focus on the regime $q\gg 1$, for simplicity. Physically, we thus have a chain of $N$ massive Dirac fermions, with relatively weaker nearest-neighbour couplings (that allow the fermions to hop from one site to the other). The Majorana particle ${R_0}$ lies at one extremity of the chain, while the  SM fermions live at the other extremity. The lightest state is (essentially) the Majorana particle $\sim {R_0}$, but it communicates to the SM through a chain of (relatively heavy, with respect to the hopping scale $m$) particles, which, as in Ref.~\cite{Giudice:2016yja}, we dub ``clockwork gears'' in the sequel.

Going to the mass eigenbasis we get a band of $N$ Dirac fermions $(\psi_{L_i} \psi_{R_i})$, $i=1 \ldots N$ , with masses
\begin{equation}
m_{\psi_i} = \sqrt{\lambda_i} \, m \ \approx \ (q-1) m \div (q+1) m\,,
\end{equation}
(the exact form of $\lambda_i$ can be found in \cite{Giudice:2016yja}) and a light Majorana field $N$
which will be our DM candidate. 
The mixing matrices are
\begin{equation}
L = U^L \psi_L\;, \qquad R = U^R \psi_R\,,
\end{equation}
with (in the limit $N\gg1$, $q\gg1$, for the general formulas see~\cite{Giudice:2016yja})
\begin{align}\label{eq:Us}
U^R_{i0} &\approx \frac{1}{q^i}\;, \qquad |U^R_{ik}| \approx \sqrt{\frac{2}{N q^2}} \times\{\pi/N \div q \}\;, \notag\\
U^L_{ik} &\approx \sqrt{\frac{2}{N}} \times\{\pi/N \div 1 \}\,.
\end{align}

In particular, the coupling of the light state with the SM is
\begin{equation}
\mathcal L \supset -\, \frac{y}{q^N} \, \bar{L}_{SM} \widetilde{H} P_R N + h.c. 
\end{equation}
which is the first key result: through the clockwork mechanism, the effective Yukawa coupling between the DM particle and SM fields is suppressed by a factor of $q^N \ggg 1$. Ultimately the value of this parameter is to be set by the requirement of stability on cosmological scales. Assuming that $m_N$ is significantly larger than the relevant thresholds, the Goldstone-boson equivalence theorem gives the decay width
\begin{equation}\label{eq:decay}
\Gamma(N \to \nu h, \nu Z, l W) \ \approx \ \frac{m_N}{8 \pi} \frac{y^2}{q^{2N}}\,.
\end{equation}
%As long as $m_N > m_W$, this only receives $\mathcal{O}(1)$ corrections, for $m_N \sim \unit[100]{GeV}$. 
For $m_N < m_W$ the decay processes are at 1-loop level, and the bound on $q$ and $N$ below is slightly weaker. By requiring that the decays \eqref{eq:decay} are slower than the age of the Universe $\tau_{AU} \simeq 4 \times 10^{17} s$, we get the bound:
\begin{equation}
q^{2N} > 6 \times 10^{41} \, \left(\frac{m_N}{\unit{GeV}}\right) y^2\,.
\end{equation}
By requiring that the lifetime is larger than the typical indirect-detection bound $\tau \sim 10^{26} s$, we get instead
\begin{equation}
\label{eq:bound}
q^{2N} > 1.5 \times 10^{50} \, \left(\frac{m_N}{\unit{GeV}}\right) y^2\,.
\end{equation}
As an example, for $m_N \sim \unit[100]{GeV}$ and $y \sim 1$, this can be satisfied by
\begin{equation}
q \equiv{M\over m} \sim 10\;, \qquad N \sim 26\,,
\end{equation}
hence for a mild hierarchy of mass scales $m$ and $M$. For smaller values of $q$, one needs, for instance, 
\begin{equation}
q  \sim 1.5\;, \qquad N \sim 150\,.
\end{equation}
Note that this conclusion is not ruined by loop corrections. The corrections to the decay width of $N$ in Eq.~\eqref{eq:decay}  are all suppressed by the clockwork mechanism, because the fermion line starting from $N$ has to go through the whole clockwork chain to mix with $R_N$ and interact with the SM fermions. Equivalently, an entire chain of spurions breaking the individual symmetries in \eqref{eq:group} is needed, in order to link the DM particle to the SM, thus giving rise to the usual clockwork suppression by a factor $q^N$.

So far the Lagrangian of Eq.~(\ref{eq:lag1}) leads to a long lived DM candidate but does not provide any DM annihilation process, apart from $N N\rightarrow L\bar{L}$ and $NN \rightarrow H H^\dagger$, which are doubly suppressed by the clockwork chain and thus far too slow to account for the DM relic density. Therefore a clockwork WIMP DM scenario requires extra interaction(s) that are not suppressed by the clockwork chain. In the following we will consider two simple possibilities. 

\paragraph*{\bf Scenario 1.} A first one consists in simply promoting the spurions into dynamical scalar fields, acquiring a vev so that (again, taking them universal for simplicity) the $m$ and $M$ parameters above are now defined as
 \begin{equation}
y_S \langle S_i \rangle = m \;, \qquad y_C \langle C_i \rangle  = M \equiv q\,m  \quad {\rm with}\quad  q > 1\,.
\end{equation}
In this case Eq.~(\ref{eq:lag1}) provides Yukawa interactions and possible additional interactions are the scalar ones:
\begin{align}
\label{eq:lag1prime}
\mathcal L' &=- \sum_{i=1}^{N} \big(\lambda_S \, S_i^\dag S_i \, H^\dag H + \lambda_C \,C_i^\dag C_i \, H^\dag H \big) \notag\\
&- \sum_{i,j=1}^{N} \lambda_{SC} \, S_i^\dag S_i \,C_j^\dag C_j  \,.
\end{align}
%In the spirit of the clockwork mechanism all singlet couplings are sizeable, hence the $y_S$ and $y_C$ couplings as well as the quartic couplings of the SM doublet field to the $C_i$ and $S_i$ fields (and among the latter). As we shall see, that some of the couplings are of order one is anyway a requirement for this DM scenario. Eventually the Higgs field will be a portal to the hidden sector constituted by the clockwork chain. 
For simplicity we consider here that all the quartic couplings are of the same order (and take them to have the same value $\lambda_{C,S,CS}$ along the clockwork chain). Depending on the mass spectrum of the model,  
the quartic couplings may or may not be necessary to have an efficient annihilation channel. For instance, as will be discussed below, if $m_{S_1}<m_N$ the dominant annihilation is $DM DM \rightarrow S_1 S_1$, which proceeds even if the quartic couplings vanish. In the opposite situation the quartic couplings are necessary in order to induce $S_i$-$h$ and/or $C_i$-$h$ mixing, leading to annihilation into a pair of $h$ or into $h+S_i$ or $h+C_i$. Similarly, such quartic couplings are interesting because they lead to possibilities of direct detection.
Thus, important ingredients along this scenario 1 are the effective Yukawa couplings of $N$ with the Higgs boson $h$ and the clockwork gears. In the 4-component notation for the spinors they are given by
\begin{equation}\label{eq:coupling_psi}
\mathcal{L} \supset - \frac{\xi_j}{\sqrt{2}} \, h \, \bar{\psi}_j P_R \, N + h.c.\;, \qquad j=1\ldots N\,,
\end{equation}
where
\begin{equation}
\label{eq:effcoup}
\xi_j = y_S\sum_{i=1}^{N}  \theta_{S,i} U_{ji}^{L\dag} U_{i-1,\,0}^R \;-\; y_C \sum_{i=1}^{N} \theta_{C,i}  U_{ji}^{L\dag} U_{i 0}^R\,,
\end{equation}
having introduced the mixing parameters
\begin{equation}
S_i = \theta_{S,i} \, \frac{h}{\sqrt{2}} + \ldots \;, \qquad C_i = \theta_{C,i} \, \frac{h}{\sqrt{2}} + \ldots
\end{equation}
If  the $\theta_{S,i}$ are of similar magnitude, $\theta_{S,i} \sim \theta_{S}$, the sums in Eq.(\ref{eq:effcoup}) are dominated by the $i=1$ term for $q \gg 1$,
\begin{align}
\sum_{i=1}^{N} U_{ji}^{L\dag} U_{i-1,\,0}^R &= \mathcal{N}_0 \sqrt{\frac{2}{N+1}} \sum_{i=1}^N \sin\left( \frac{i j \pi}{N+1} \right) \frac{1}{q^{i-1}} \notag\\
& \approx \sqrt\frac{2}{N+1} \sin\left( \frac{j \pi}{N+1} \right) \\
\sum_{i=1}^{N} U_{ji}^{L\dag} U_{i 0}^R &= \mathcal{N}_0 \sqrt{\frac{2}{N+1}} \sum_{i=1}^N \sin\left( \frac{i j \pi}{N+1} \right) \frac{1}{q^{i}} \notag\\
& \approx \sqrt\frac{2}{N+1} \frac{1}{q} \sin\left( \frac{j \pi}{N+1} \right)\,.
\end{align}
Thus
\begin{align}
\xi_j \approx \sqrt\frac{2}{N+1} \sin\left( \frac{j \pi}{N+1} \right) \big( y_S \theta_{S} - \frac{1}{q} y_C \theta_{C}\big) \,,
\end{align}
where we have defined $\theta_S \equiv \theta_{S,1}$, $\theta_C \equiv \theta_{C,1}$. The important result is that the interaction of the DM particle $N$ to the Higgs boson $h$ and the clockwork gears is not suppressed by the clockwork mechanism.  In other words
\begin{equation}
\label{eq:xi}
\xi^2 \equiv \sum_{j=1}^N |\xi_j|^2 \approx  \big( y_S \theta_S - \frac{1}{q} y_C \theta_C\big)^2 \approx y_S^2 \theta_S^2 \,.
\end{equation}
For the same reason, the clockwork gears are short lived and do not constitute a thermal relic, or spoil Big-Bang Nucleosynthesis predictions.

Although these formulas are obtained from the diagonalization of the clockwork mass Lagrangian in the limit $m_N \to 0$, numerically we find that they are rather accurate as long as $m_N \lesssim 0.7-0.8 \, m_\psi$. An exact diagonalization gives that the Dirac spinors $\psi_i$ split into pseudo-Dirac Majorana spinors, and the clockwork mechanism is active as long as $m_N < (q-1) \, m \approx m_\psi$.  The exact diagonalization also gives that the light Majorana state ${N_R^c}$ acquires an overlap $\mathcal{U}^L_{i0}$ with the LH field $L_i$
\begin{equation}
\mathcal{U}^L_{i0} \approx \frac{m_N}{m\, q^{i+1}}\,,
\end{equation}
whereas the $\mathcal{U}^R_{i0}$ overlap between $N_R$ and the $R_i$ field is approximately as $U^R_{i0}$ in \eqref{eq:Us}. This leads to the couplings
\begin{align}
- \mathcal L \supset \frac{m_N}{m q^2} \, \overline{N^c} N \big( y_S S_1 - \frac{1}{q} y_C C_1\big) + h.c.
\end{align}
and, in particular, to the effective coupling to the Higgs boson, which is relevant for direct detection:
\begin{align}\label{eq:coupling_N}
- \mathcal L &\supset \frac{m_N}{m q^2} \, \overline{N^c} N \frac{h}{\sqrt{2}} \big( y_S \theta_S - \frac{1}{q} y_C \theta_C\big) + h.c. \notag\\
&\approx \frac{m_N}{\sqrt{2} m q^2} \xi\, \overline{N^c} N h  + h.c.
\end{align}

Note that, whereas the exact amount of clockwork suppression and the precise masses of the states in the band depend on the details of the couplings along the whole clockwork chain (which, as said above, we assumed to be universal for simplicity),  for $q \gg 1$ the couplings relevant for DM freeze-out and direct detection essentially depend only on the coupling of $N$ with the first clockwork chain state, $L_1$, therefore being quite insensitive to the exact profile of the clockwork.

\paragraph*{\bf Scenario 2.} The second possibility 
%of interactions inducing 
for annihilations that we will consider is from a single scalar field $\phi_N$ to couple directly with $R_0$ (with all $S_i$ and $C_i$ being non-dynamical)
\begin{equation}
\label{eq:Lscenario2}
\mathcal L' =- \frac{y_N}{2} \phi_N \overline{R_0^c} R_0 + h.c.-\lambda_N \phi_N^\dagger \phi_N H^\dagger H\,,
\end{equation}
whose vev induces both the $R_0$ soft mass of Eq.~(\ref{eq:lag1}), $m_N= y_N \langle \phi_N \rangle$, and mixing of $\phi_N$ with the Higgs boson (through the $\lambda_N$ interaction). For what concerns the DM annihilation, this scenario is similar to a standard Majorana DM Higgs-portal scenario. The annihilation may proceed through $NN\rightarrow \phi_N \phi_N$ or into SM fields through $h$-$\phi_N$ mixing.
%, with the $\phi_N$ decaying into SM fields through the mixing too.

%%%%%%%%%%%%%%%%%%%%%%%%%%%%%%%%%%%%%%%%%%%%%%%
\section{Dark matter phenomenology}
\label{sec:pheno}
In this section, we discuss the basic phenomenology of the above DM model, in particular its abundance set by thermal freeze-out and the constraints from direct detection through the Higgs coupling. Some comments on collider limits are discussed at the end of this section. The relic abundance depends essentially on the relative hierarchy of masses between DM  and the scalars $h$ and $S_i$ (we assume in the sequel that $m_h < m_{S_i}$). Here and in what follows, with a slight abuse of notation, $S_i$ denote the (canonically normalized) real part of the respective complex fields. 

\begin{figure}
\includegraphics[width=0.40\textwidth]{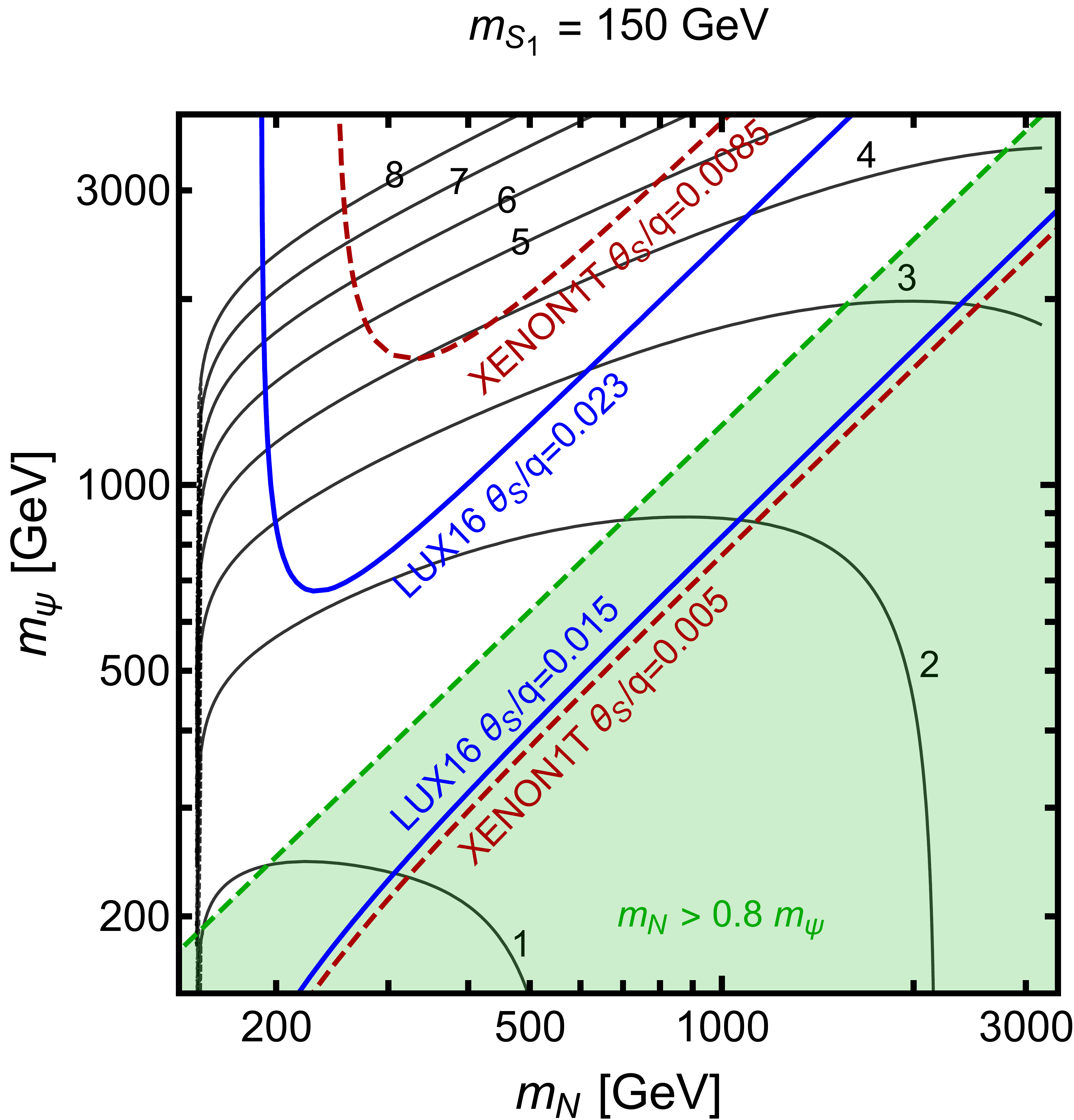}
\caption{The coupling $y_S$ required to obtain the observed relic density in Scenario 1A, for $m_{S_1} = \unit[150]{GeV}$ and $q \gg 1$. We also show direct-detection limits (blue, continuous) and future prospects (red, dashed), from LUX 2016 and XENON1T, respectively. For a fixed value of $\theta_S/q$, the excluded (probed) region is below the blue (red) lines.\label{fig:yS_II}}
\end{figure}

\medskip
\subsection*{Scenario 1A: $m_{S_1} < m_N$}
\label{subsec:secluded}
In this regime,  the dominant channel that drives the relic abundance is annihilation into the scalar $S_1$ 
\begin{equation}
NN \to S_1 S_1\,,
\end{equation}
with t- and u-channel exchange of the $\psi_i$, the same ones involved in the DM decay. Even if kinematically allowed, all other channels (into other $S$  or $C$ states) are suppressed by extra powers of $1/q$. The relevant coupling of $S_1$ is
\begin{equation}\label{eq:coupling_s1}
\mathcal{L} \supset - \frac{\xi^S_j}{\sqrt{2}} \, S_1 \, \bar{\psi}_j P_R \, N + h.c.\;, \qquad j=1\ldots N\,,
\end{equation}
with 
\begin{equation}
\sum_{j=1}^N |\xi^S_j|^2   \approx y_S^2 \,,
\end{equation}
and the cross-section is
\begin{align}\label{eq:cross_section}
\sigma v &= \frac{y_S^4 v^2}{384 \pi} \; \frac{m_N^2}{(m_\psi^2 + m_N^2 - m_{S_1}^2)^4} \, \sqrt{1- \frac{m_{S_1}^2}{m_N^2}} \notag \\
& \times \; \left( 3 m_\psi^4 + 2 m_\psi^2 (m_N^2 - m_{S_1}^2) + 3 (m_N^2 - m_{S_1}^2)^2\right)\,,
\end{align}
having approximated $m_{\psi_j} = m_{\psi} \equiv q m$. Annihilation is p-wave suppressed, as expected for a pair of Majorana DM particles annihilating into scalars. 

The effective spin-independent (SI) coupling to nucleons is determined by \eqref{eq:coupling_N} and is found to be (using the Higgs-nucleon coupling of~\cite{Cline:2013gha})
\begin{equation}
G_{SI} = \frac{2 m_N \xi}{\sqrt{2} m q^2} \, \frac{f_N m_{nuc}}{v\, m_h^2}\,,
\end{equation}
with $\xi$ as given in Eq.~(\ref{eq:xi}), $f_N = 0.30$, $m_{nuc} = \unit[0.946]{GeV}$
and the direct-detection nucleon cross-section is
\begin{equation}\label{eq:dd}
\sigma_{SI} = \frac{1}{\pi} \frac{m_N^2 m_{nuc}^2}{(m_N+m_{nuc})^2} G_{SI}^2\,.
\end{equation}

In Fig.~\ref{fig:yS_II} we plot the coupling $y_S$ (solid black lines) required for the relic density, for $m_S= \unit[150]{GeV}$ and $q \gg 1$ (so that one is in the regime of validity of \eqref{eq:cross_section} and \eqref{eq:dd}). 
In the region $m_N \gtrsim 0.8 \,m_\psi$, as discussed above, the relations given in the previous section become less accurate, whereas the clockwork mechanism stops working altogether for $m_N \approx m_\psi$ (hence the green/shaded region is excluded). Moreover, the plot should be considered valid as long as the phase-space is sufficiently open so that processes other than $N N \rightarrow S_1 S_1 $  may be neglected for determining the relic abundance of DM. 
Direct detection is excluding regions below the blue lines, from LUX 2016~\cite{Akerib:2016vxi}. Future prospects for XENON1T~\cite{Aprile:2015uzo} are also given, this probing the regions below the red dashed lines. One observes that this scenario is compatible with the perturbativity bound $y_S < \sqrt{4 \pi} \approx 3.5$ for $m_N,\,m_{\psi} < \unit[2]{TeV}$. For the benchmark model parameters of Fig.~\ref{fig:yS_II}, XENON1T will test completely the allowed region, for $\theta_S/q > 0.0085$. 
Note also that in the presence of quartic couplings, the $S$ and $C$ states decay into SM fields via mixing with $h$ and thus do not provide any relic on top of the DM.

\begin{figure}
\includegraphics[width=0.40\textwidth]{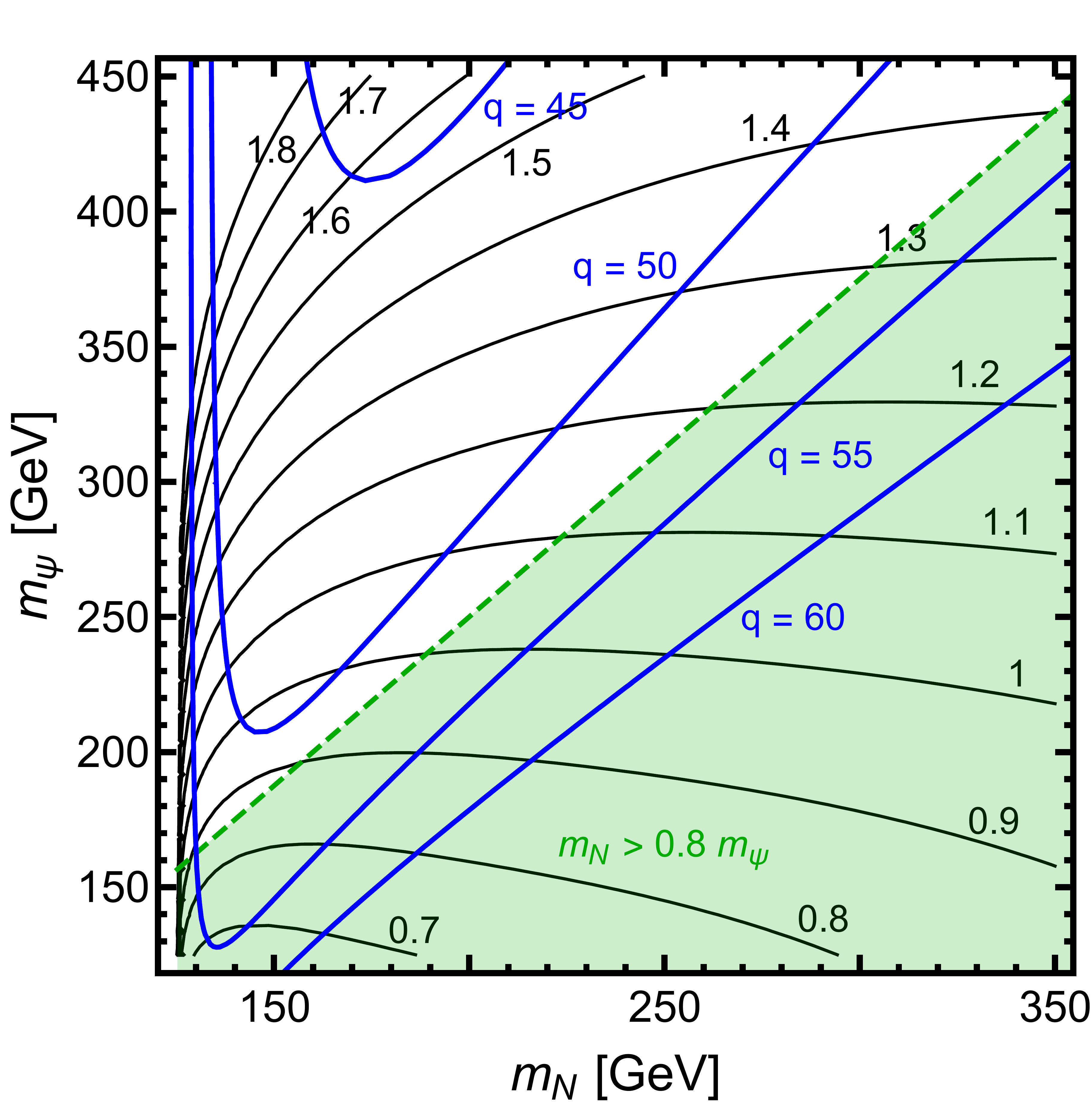}
\caption{The effective coupling $\xi = \theta_S y_S$ required to obtain the observed relic density in Scenario 1B. Direct-detection limits from LUX 2016 are shown in blue. 
%For each fixed value of $q$, the allowed parameter region is above the blue line
\label{fig:NNhh}}
\end{figure}

\subsection*{Scenario 1B: $m_N < m_{S_i}$ and $2m_N < m_{S_i} +m_h$}
When $m_{N} > m_h$, we are far from the Higgs resonance and the relic density is mainly determined by the process
\begin{equation}
N N \to h h\,,
\end{equation}
with t- and u-channel exchange of $\psi_i$. The relevant couplings are given in \eqref{eq:coupling_psi} and the cross-section can be obtained from \eqref{eq:cross_section} with the replacements
\begin{equation}
m_{S_1} \to m_{h}\,, \qquad y_S \to \xi = y_S \theta_S\,.
\end{equation}
This
%, together with the mass of $h$ 
is the main difference with respect to scenario 1A: the effective coupling $\xi$ is now suppressed by mixing with the SM Higgs. The direct-detection cross section is through the Higgs boson, as in scenario 1A, and so is as in Eq.~\eqref{eq:dd}.

In Fig.~\ref{fig:NNhh} we show in the $m_N-m_\psi$ plane the effective coupling $\xi$ (black solid lines) required to have the observed relic density, as well as the direct-detection exclusion limit from LUX 2016. Again, in the green region $m_N \gtrsim 0.8 \,m_\psi$ our approximate expressions become less accurate, while the clockwork mechanism is active up to $m_N \approx m_\psi$. Direct-detection exclusion limits require a large  value of $q$. As Fig.~\ref{fig:NNhh} reveals, for $q \gtrsim 50$ or so the parameter space above the line $m_N \sim 0.8\,m_\psi$ is {\em a priori } acceptable. Now, relic abundance requires that the effective coupling  is $\xi \gtrsim 0.8$. For $\unit[130]{GeV} < m_S < \unit[300]{GeV}$ the phenomenological bound on the mixing of a \emph{single} scalar singlet with the Higgs boson is  \cite{Falkowski:2015iwa,Robens:2015gla,Chalons:2016jeu} $\theta \lesssim 0.3-0.4$.
By assuming universal mixing of all the $S_i$ with $h$, this implies
\begin{equation}
\theta_S \lesssim \frac{0.4}{\sqrt{N}} \,,
\end{equation}
with $N \gtrsim 16$ to account for the lifetime of the DM particle, Eq.(\ref{eq:bound}).
Thus $\xi > 0.8$ corresponds to non-perturbative values of $y_S$.
Therefore, for $m_N > m_h$, scenario 1B does not work, unless one gives up the assumption of universality and the mixing $\theta_{S,1}$ is bigger than the average of $\theta_{S,i}$. For instance, by taking $\theta_{S,1} = 0.3\, (0.4)$, perturbativity requires $\xi \lesssim 0.3\, (0.4) \times \sqrt{4\pi} \simeq  1.1 (1.4)$, which is compatible with all the constraints, for $m_N \lesssim \unit[225\, (350)]{GeV}$, $m_\psi \lesssim \unit[275\, (425)]{GeV}$. We see that this scenario has a smaller parameter space available, as compared to scenario 1A, 
due to the extra Higgs boson mixing suppression,
% since the coupling relevant for the relic density is now suppressed by the mixing with the Higgs boson, 
which in scenario 1A only affects direct detection. 

Alternatively, this scenario works near the Higgs resonance $m_N \approx m_h/2$, also for universal mixing. To see this, notice that in this regime the relevant coupling for both the relic density and direct detection is \eqref{eq:coupling_N}, {\em i.e.} essentially the one of a Higgs-portal Majorana DM. By adapting the results of \cite{Beniwal:2015sdl}, by means of the replacement 
\begin{equation}
\frac{\lambda_{h\chi}}{\Lambda_\chi} \ \to \ \frac{4 m_N}{\sqrt{2} v q m_{\psi}}\, \xi \;,
\end{equation}
the results of \cite{Beniwal:2015sdl} give $\unit[55]{GeV} < m_N < \unit[62.5]{GeV}$ and
\begin{equation}
\xi \simeq 0.14 \times \frac{q}{10} \, \frac{m_{\psi}}{\unit[100]{GeV}} \, \times \, c_{\rm res}\,,
\end{equation}
where $c_{\rm res}$ varies between 1 (for $m_N \simeq \unit[55]{GeV}$) and 0.2 (for $m_N \simeq \unit[62]{GeV}$).

Finally, this scenario also works near the $S_1$ resonances, {\em i.e.}~for $2 m_N \approx m_{S_i}$ but, since in this case the phenomenology depends strongly on the unknown values of $m_{S_i}$, this option will not be discussed further here.

\begin{figure}
\includegraphics[width=0.40\textwidth]{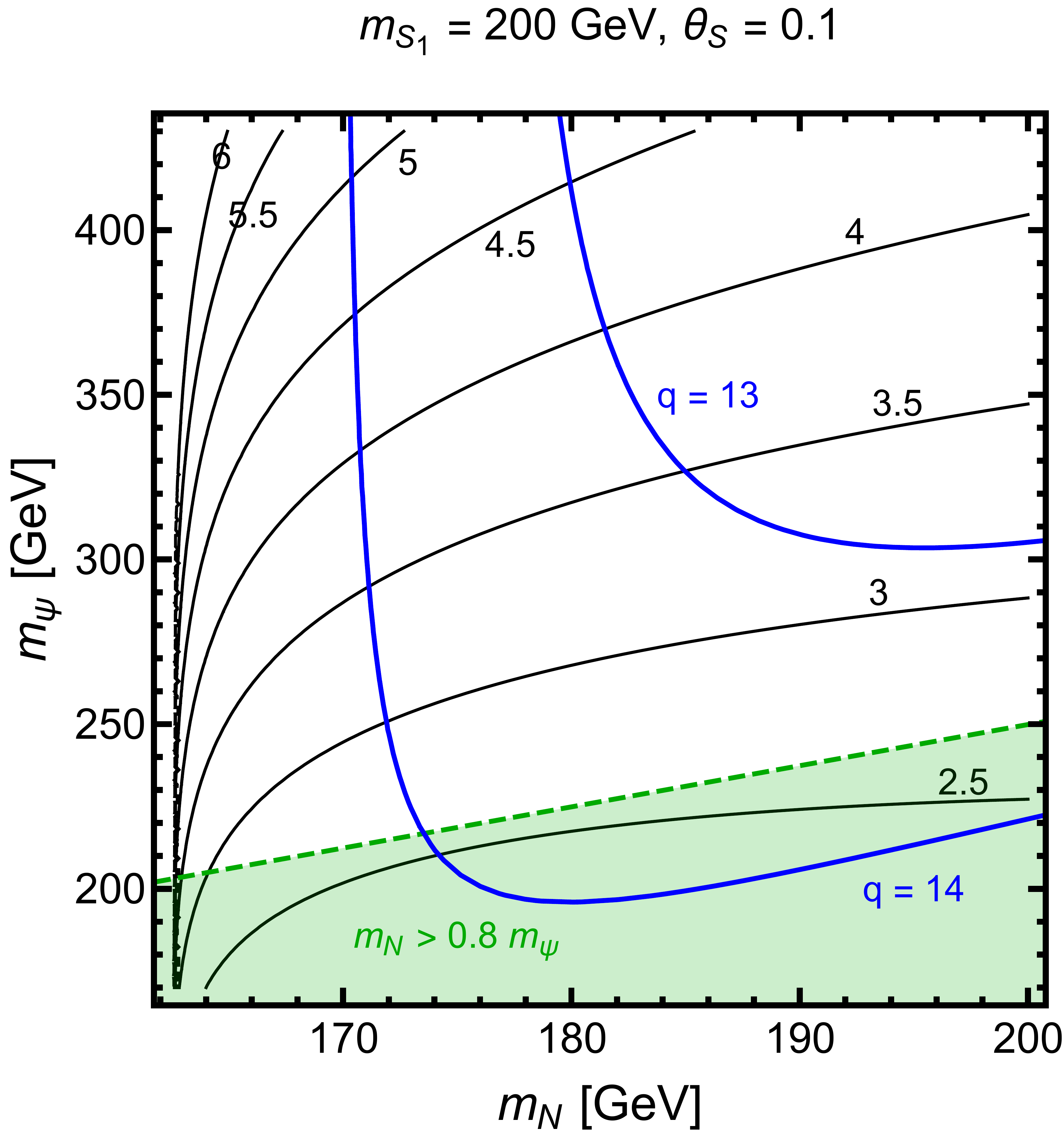}
\caption{The coupling $y_S$ required to obtain the observed relic density in Scenario 1C, for $m_{S_1} = \unit[200]{GeV}$, $\theta_S=0.1$. Direct-detection limits from LUX 2016 are shown in blue.\label{fig:yS_III}}
\end{figure}

\medskip
\subsection*{Scenario 1C: $m_{S_1} + m_h < 2 m_N < 2 m_{S_1}$}
The dominant channel for relic density is:
\begin{equation}
NN \to S_1 h\,.
\end{equation}
The cross-section has a more complicated form and will not be given here. Since $\sigma v \propto y_S^4 \theta_S^2$, this scenario interpolates between the previous two, away from the resonances.

For universal mixing of all the $S_i$ with $h$, we find that only $m_N < \unit[240]{GeV}$ (with $m_{S_1} < \unit[300]{GeV}$) is compatible with perturbativity, although in a narrow region of the parameter space. Relaxing the universality assumption, the parameter space opens up. For instance, in Fig.~\ref{fig:yS_III} we show the results for $\theta_S=0.1$, $m_{S_1} = \unit[200]{GeV}$. We see that in this case direct-detection constraints require $q \gtrapprox 13$, which is a significantly milder constraint, with respect to Scenario 1B.

\bigskip
\subsection*{Scenario 2: non-dynamical spurions}
If the spurions $m$ and $q m$ in \eqref{eq:clock_lagr} do not have dynamical origin (or the fields $S_i$ and $C_i$ are decoupled), one can still have a successful channel for the annihilation of the DM if, for instance, $m_N$ has a dynamical origin, being generated by the vev of a scalar $\phi_N$, as given in Eq.~(\ref{eq:Lscenario2}). For heavy $\phi_N$, this scenario effectively reduces to a Higgs-portal model, which works only near the Higgs resonance \cite{Robens:2015gla}. For light $\phi_N$, this scenario can work, for instance, when the annihilation into two $\phi_N$ is kinematically open. However, we will not discuss this scenario in detail, since the relevant phenomenology is decoupled from the clockwork mechanism. In particular, in scenario 2 the clockwork chains $\psi$ do not need to be at reach of current or near-future experiments, even if this is a perfectly viable possibility.

\medskip

\subsection*{Other existing limits and future prospects}
In the Scenario 1 discussed above, the clockwork gears $\psi_i$ need to be at \unit{TeV}  or lighter. In particular, in Scenario 1B and 1C, the combination of perturbativity considerations and the existing limits on Higgs-singlet mixing requires $m_\psi$ to be at most at a few hundreds \unit{GeV}, whereas in Scenario 1A the clockwork gears can be as heavy as $\sim$~\unit{TeV}, if one allows large values of $y_S$ at the boundary of the perturbativity region. On the other hand, in the philosophy of the clockwork mechanism the coupling to the SM doublet $y$ should not be too small either. Therefore, a natural consequence of the framework discussed in this note is the presence of observable effects of the states $\psi_i$. The coupling of the mass eigenstates $\psi_i$ to $L_{SM}$ proceeds via their (large) overlap with the clockwork state $R_N$. Thus, as long as $q \gg 1$, we may reason in terms of a single state with mass $m_\psi$ and Yukawa coupling $y$ to the SM. If the relevant experiments are instead able to resolve the different eigenstates $\psi_i$, one should think in terms of a collection of $N$ states with couplings $\sim y/\sqrt{N}$. The intermediate regime, in which the effective description of the band of gears in terms of a single state is not valid, but at the same time the gears are not well separated, is much more involved and a detailed analysis goes beyond the purposes of this work.

Phenomenologically, the states $\psi_i$ are essentially a collection of pseudo-Dirac heavy neutrinos. Their collider signatures are studied in detail \cite{Das:2014jxa}. The most stringent existing constraint come from electroweak precision tests (EWPT) and gives
\begin{equation}
|B_{l\psi}|^2 \ \equiv \ \frac{v^2}{2} \frac{|y_l|^2}{m_{\psi}^2} \ \lessapprox \ 10^{-3}\,,
\end{equation}
for the muon flavour, and slightly weaker for the electron flavour. The most sensitive channel for their direct production at LHC is the trilepton plus missing energy~\cite{Das:2014jxa}. In particular, the $\sqrt{s} = \unit[14]{TeV}$ data with $\unit[300]{fb^{-1}}$ integrated luminosity will probe couplings allowed by EWPT up to $m_\psi \approx \unit[200]{GeV}$, which is the region required in Scenarios 1B and 1C.

A stronger indirect bound is provided by LFV in processes like $\mu \to e \gamma$ and $\mu \to e$ conversion in nuclei. The final MEG limit \cite{TheMEG:2016wtm} on the branching ratio for $\mu \to e \gamma$ is given by
\begin{equation}
BR(\mu \to e \gamma) \ \approx \ 8 \times 10^{-4} \, |B_{e\Psi}|^2 \, |B_{\mu\Psi}|^2 < 4.2 \times 10^{-13}\,.
\end{equation}
Although this limit is stronger than the one from EWPT for $|B_{e\psi}| \simeq |B_{\mu\psi}|$, it crucially depends on both the muon and electron couplings being sizeable. This result will be improved in the near future by 4 orders of magnitudes by the $\mu \to e$ conversion experiments  Mu2e~\cite{mu2e} and COMET~\cite{comet}. Thus, under the reasonable assumption of not having a large hierarchy between the muon and electron couplings, LFV provides the best prospects to probe the framework in the near future.

If $m_N$ is not much smaller than $m_\psi$, the pseudo-Dirac states split into Majorana pairs with different masses and couplings to the SM leptons. In this case, the most sensitive additional channel is provided by same-sign dilepton plus jets with no missing energy. The $\unit[300]{fb^{-1}}$ dataset will allow to probe couplings allowed by EWPT up to $m_\psi \approx \unit[300]{GeV}$~\cite{Deppisch:2015qwa}. Finally, future electroweak precision tests at lepton colliders would provide a strong indirect probe of the mechanism.

As for the scalar sector, Scenario 1B and 1C require a large mixing between the scalar $S_1$ and the Higgs boson, whereas in Scenario 1A (2) the mixing of $h$ with $S_1$ ($\phi_N$) can be either large or small. A detailed analysis of the relevant constraints is given in~\cite{Falkowski:2015iwa,Robens:2015gla,Chalons:2016jeu}. For $m_{S_1} < \unit[500]{GeV}$ the most stringent constraint comes from the direct searches of Higgs-like resonances, giving $\theta_S < 0.3-0.4$. For higher masses the most stringent limit is again given by EWPT: $\theta_S \lessapprox 0.3$. Therefore, the scalar sector of our framework can give observable signatures at Higgs searches, for $m_{S_1}$ of few hundreds \unit{GeV}, or at future electroweak precision measurements, for larger $m_{S_1}$.

We conclude this section with a brief comment regarding indirect detection. In the scenarios discussed here, the annihilation of DM is p-wave suppressed, and thus it is  relevant only at the time of thermal freeze-out. However, through the clockwork gears, the DM particle can decay on a time scale that may be 
%very long but yet may be 
relevant for indirect searches, $\tau \sim 10^{26}$ sec. As the decay is through the coupling with the SM Higgs and leptons, the most notable potential spectral feature is 
%that the decay products will include
the presence of mono-energetic neutrinos from the DM decay into $h+\nu$, a signature that has been studied in the literature (see \cite{Aisati:2015vma} and references therein).

%has been already much studied in the literature and is promising . 

\medskip
\section{Clockwork WIMP from deconstruction} \label{sec:deconstr}

We now want to investigate whether the scenario we advocate can be related to the deconstruction of an extra dimension \cite{Giudice:2016yja} in a simple way. In this section we show how  this can be achieved in a \emph{flat} spacetime for a fermionic clockwork chain, differently from the construction in~\cite{Giudice:2016yja}, which makes use of a curved metric.

We start from the Lagrangian of a free massive fermion in 5D
\begin{equation}
\label{eq:lag2}
{\cal L}_5  \supset \bar \psi ( i \! \overleftrightarrow{\slash\!\!\!\partial}_{\!\!\!D} - M) \psi \equiv i \bar \psi \gamma^\mu \partial_\mu \psi + i \bar \psi \gamma^Z  \overleftrightarrow{\partial}_{\!\!\!Z} \psi - M \bar \psi \psi\,,
\end{equation}
where $Z$ is the coordinate in the extra dimension, compactified with length $\pi R$,  $\gamma^Z \equiv i \gamma_5$, and $\overleftrightarrow\partial \equiv (\overrightarrow{\partial} - \overleftarrow{\partial})/2$. In the Weyl basis, $\psi = (L \; R)^{\rm T}$ and we get
\begin{align}\label{eq:lagr5D}
\mathcal{L}_5 \supset i \overline{\psi} \gamma^\mu \partial_\mu \psi \, + \, \bigg[\frac{1}{2} \Big( \overline{L} \partial_Z R - (\partial_Z \overline{L}) R \Big) \, - M \overline{L} R\, + h.c.  \bigg]\,.
\end{align}
In  a theory with fermions, it is well known that a naive discretization gives rise to the famous doubling problem on the lattice (see {\em e.g.}~\cite{Creutz:1984mg}), which may be removed using a Wilson term (see the discussion in \cite{Giudice:2016yja}):
\begin{equation}\label{eq:Wilson}
\mathcal L \supset - \frac{a}{2} \, \partial_Z \overline{\psi} \, \partial_Z \psi = - \frac{a}{2} \, \partial_Z \overline{L} \, \partial_Z R \, +h.c. 
\end{equation}
where $a = \pi R/N \to 0$ is the lattice spacing in the extra dimension. This term, which vanishes in the continuum limit, guarantees that the discrete theory tends to the continuum one~\eqref{eq:lag2} in the appropriate limit $N \to \infty$.
By discretizing \eqref{eq:lagr5D} and \eqref{eq:Wilson} along the extra dimension and redefining conveniently the fields\footnote{To obtain a form that matches exactly \eqref{eq:clock_lagr}, we have added boundary counterterms, following~\cite{Giudice:2016yja}. However, this is not necessary for the clockwork mechanism to work, but only for convenience of the presentation.} we find
\begin{align}
\mathcal{L} &\supset \sum_{i=1}^{N} \; i \overline{\psi}_i \gamma^\mu \partial_\mu \psi_i \; + \; \sum_{i=1}^{N-1} \frac{1}{a} \, \Big( \overline{L}_i R_{i+1} \,+ h.c.\Big) \notag \\
&- \sum_{i=1}^{N} \bigg[ \bigg( \frac{1}{a} + M \bigg) \, \overline{L}_i R_i \, + h.c. \bigg]\,.
\end{align}
The last ingredient to obtain the clockwork chain as in~\eqref{eq:clock_lagr} is the addition of a chiral field $R_0$ localized at $Z = 0$, with a Majorana mass $m_N$, interacting with the bulk fermion $\psi$, whereas the SM is localized at $Z= \pi R$. Notice that the precise value of the coupling of the interaction term $ \overline{L} R_0 \delta(Z)$ does not affect significantly the clockwork mechanism since, although the overall clockwork suppression is proportional to this coupling, the exponential part of the suppression does not depend on it. Equivalently, one can instead impose the Dirichlet boundary condition $L(0)=0$ which effectively removes $L_0$ and leaves one unpaired right-chiral mode $R_0$ in the discretized theory. We thus end up with the clockwork Lagrangian in \eqref{eq:clock_lagr} with the identifications
\begin{equation}\label{eq:ident_5D}
m \equiv \frac{1}{a} \;, \qquad q m \equiv \frac{1}{a} + M\,.
\end{equation}
In the $N \to \infty$ we obtain a \emph{finite} clockwork suppression:
\begin{equation}
q^N = \bigg( 1 + \frac{\pi R M}{N} \bigg)^N \ \to \ e^{\pi R M}\,.
\end{equation}

The spectrum of the clockwork states in the continuum limit consists of the DM, with a mass $\sim m_N$ and a tower of Kaluza-Klein modes (the clockwork gears) starting at the mass $M$. Since, for the effectiveness of the clockwork, $M \gg 1/R$, the Kaluza-Klein modes will have mass differences $\Delta M \sim 1/R \ll M$, \emph{i.e.}~they will not be largely separated one from each other.

The clockwork setup that we envision superficially resembles that of domain wall fermions (see {\em e.g.} \cite{Kaplan:2009yg}), with a chiral mode localized on one wall, and the (chiral) SM degrees of freedom localized on another wall. The correspondance is only approximate because we assume only one chiral mode, strictly located at the $Z=0$ wall, which furthermore must have a Majorana mass. 

As for realizing the dynamical scalar fields needed for DM annihilation, we distinguish between the two scenarios discussed in the previous sections.

\paragraph*{\bf Scenario 1.}
One possibility is to add a scalar $\phi$ in the bulk, with a Yukawa interaction with $\psi$:
\begin{equation}
\mathcal{L}_5 \supset y_\phi \, \overline{\psi} \psi \phi\,,
\end{equation}
which can also generate the Dirac bulk mass $M$, if it acquires a vev. 
After discretization, this will give rise, effectively, to the interaction with the fields $C_i$ discussed in section \ref{sec:clock}. 
Thus the $C_i$ are dynamical fields, unlike the $S_i$ which all come from the discretization of the fifth dimension, $S_i=1/a$.
This framework predicts the universality of the clockwork chain.
In this case obviously the annihilation cannot proceed into $S_i$ states but it can into $C_i$ states, just in the same way as considered in sections~\ref{sec:clock} and \ref{sec:pheno}, replacing $S_1$ by $C_1$ and $y_S$ by $y_C/q$ in Eqs.~(\ref{eq:coupling_s1})-(\ref{eq:cross_section}). The scenario is perfectly viable provided that $q$ is not too large. 
%Similarly to what we did in Fig.~1 for $DM DM \rightarrow S_1 S_1$, Fig.~? gives the constraints if the annihilation proceed from $DM DM \rightarrow C_1 C_1$.
Note that in the limit $N \to \infty$, \emph{i.e.~}$q \simeq 1$, these $C_i$ interactions are not suppressed with respect to the ones of $S_i$ in Eq.~(\ref{eq:coupling_s1}) and in this case all the $DM \, DM\rightarrow C_i C_i$ processes are relevant. If the bulk mass term $M$ is generated by the vev of $\phi$, Eq.~\eqref{eq:ident_5D} gives
\begin{equation}
q m \equiv  \frac{1}{a} + y_\phi \langle \phi \rangle = m + y_\phi \langle \phi \rangle\,.
\end{equation}
As for the quartic couplings of the $C_i$ and the Higgs boson in \eqref{eq:lag1prime}, these can be sizeable if the field $\phi$ has a sizeable overlap with the brane at $Z= \pi R$, or if the Higgs field is not confined to this brane.\footnote{In the latter case, the dimension-5 scalar doublet leads to $N$ copies of the Higgs doublet at the different sites in theory space. At this stage, we may simply assume that the SM Higgs is the lightest mass eigenstate. }

Another  possibility is to gauge the fermionic field $\psi$ with an abelian group, which gives rise to couplings of the form
\begin{equation}
\overline{L}_{i} U_i R_{i+1} \qquad \mbox{\rm with} \qquad U_i = \exp(i  A_i )\,,
\end{equation}
where $ A_i$ is the component of the gauge field (technically, averaged along the link between sites $i$ and $i+1$ times the lattice spacing $a$). Under a gauge transformation, $\psi_i\rightarrow e^{i  \theta_i} \psi_i$, $A_i \rightarrow  A_i + (\theta_{i+i} - \theta_i)$. This is of course well known. The relevant point is that here the $C_i$ are spurions generated by the $M$ mass term and the link variables $U_i$ play the role of the fields $S_i$ in section~\ref{sec:clock}. The relevant $U_i$ degrees of freedom are nevertheless not real fields, like instead the $S_i$ considered in Eqs.~(\ref{eq:coupling_s1})-(\ref{eq:cross_section}), but phases, which from the $4D$ perspective are pseudo-scalar fields.
%, instead of the real scalars that we have considered in the previous sections. 
In this case one expects that their mixing with the Higgs boson is at least suppressed, and therefore only scenario IA would work (in the same way as in section~\ref{sec:pheno}).

Finally, notice that in the case of a curved spacetime along the extra dimension, as considered in \cite{Giudice:2016yja}, with a metric of the form $d s^2 = e^{\frac{4}{3} k |Z|} (d x^2 + d Z^2)$ originating from a model with a dilaton $\mathcal S$, a similar scenario could also be realized, provided that the UV completion of the model fulfills certain conditions. For instance, in order to obtain a Yukawa interaction with a bulk scalar $\phi$ not exponentially-suppressed for the DM light mode, one has to choose conveniently the spurion charge under dilatations of the (dimensionful) bulk Yukawa coupling. In particular, one can show that the unique successful choice is to take $y_\phi \to e^{\alpha} y_\phi$, under the global Weyl transformation\cite{Giudice:2016yja} $g_{MN} \to e^{-2 \alpha} g_{MN}$, $\mathcal S \to \mathcal S +  3 \alpha$. With this choice, the warping coming from the metric and the interaction with the dilaton combine to give an unsuppressed interaction with the DM mode, localized at $Z \simeq \pi R$ in the construction of~\cite{Giudice:2016yja}.

% $\sqrt{-g} = e^{-8 k Z/3}$, the bulk gauge and Yukawa couplings considered above would be suppressed, for the light eigenmode, by a factor $e^{- 4 k \pi R/3}$, when written in terms of the canonically-normalized 4D fields in Minkowski spacetime. Thus, the WIMP paradigm could not be realized (at least in simple terms) in Scenario 1, having adopted the curved-space construction of \cite{Giudice:2016yja}.

\paragraph*{\bf Scenario 2.}
In this case one only needs a scalar field localized on the brane $Z=0$, with a ``Majorana'' interaction with $R_0$:
\begin{equation}
\mathcal{L}_5 \supset \phi_N \, \overline{R_0^c} R_0 \, \delta(Z) \,.
\end{equation}
To generate the quartic scalar coupling $\lambda_N$ in \eqref{eq:Lscenario2}, which need to be nonzero (to provide decay channels of $\phi_N$ into the SM) but not necessarily sizeable, one may assume that another (free) scalar field is present in the bulk, with a significant overlap with both the branes. In this case the structure consists of two branes where interactions occur, connected by free fields in the bulk. In alternative, one may take the Higgs field as living in the bulk and no extra bulk scalar.

Finally, notice that this scenario 2 would also work in the extra-dimensional construction of~\cite{Giudice:2016yja} because the interaction is essentially 4D, and therefore not exponentially suppressed by the metric, provided that a decay channel of $\phi_N$ into the SM is present.

\medskip
\section{Majorana neutrino masses} \label{sec:neutrino}

As the SM left-handed neutrinos have large Yukawa couplings with the TeV-scale clockwork gears $\psi_i$, one could worry that these couplings induce far too large SM neutrino masses. However, this is not the case because, if there were no $R_0$ at the other end of the clockwork chain, the left-handed neutrinos would have no chiral partner to get a mass from. Since  the interactions of $R_0$ with the SM fermions have to proceed along the whole clockwork chain, the light-neutrino masses receive a suppression from the clockwork mechanism
\begin{equation}
\label{eq:mnu}
m_{\nu} \simeq \frac{m_D^2}{q^{2N} m_N}\,,
\end{equation} 
where $m_D= y v$, and too large neutrino masses are not generated. 
On the contrary, since the suppression needed for the DM lifetime is much larger than for the neutrino masses ({\em i.e.}~typically, in the ordinary language of effective operators,  a dimension-6 versus dimension-5 GUT scale suppression) the mass induced in the above framework is far below the masses needed: $m_\nu\sim 10^{-40}$~eV for $m_N\sim 100$~GeV.

However, it is worth to point out that a fermion clockwork chain  can induce a Majorana neutrino mass in agreement with data through Eq.~(\ref{eq:mnu}) if this chain involves a smaller $q^{2N}$ factor. For instance for $q=10$, $m_\nu\sim 10^{-1}$~eV requires $2 N\sim 15 - log_{10}(m_N/\unit{GeV})$.  
%In this case the $R_0$ cannot be a DM candidate because it would have a far too short lifetime: $\tau \sim \unit[10^{-12}]{s}$, for $m_N = \unit[100]{GeV}$. \ToH{leptogenesis}. 
Alternatively if a $R_0$ has no Majorana mass, a Dirac mass is induced as in \cite{Giudice:2016yja}, which for $q=10$ requires $N\sim 12$.
As one could have expected from the start, the Dirac case needs more clockwork suppression than the Majorana case. 

Experimentally we know that there are at least two nonzero neutrino masses, and to generate at least two of them there are essentially two simple options, as we discuss now.
\paragraph*{\bf Option 1.} If one assumes that only the last site ({\em i.e.}~$R_N$) couples to the SM leptons, in both the Majorana and Dirac cases a single clockwork chain is not enough, because it can only induce one neutrino mass, given that the chain couples to a single combination of $L_{e,\mu,\tau}$. In order to induce 2 (3) neutrino masses one therefore needs 2 (3) clockwork chains, including a different $\nu$ chiral partner for each chain.
%Thus, one could imagine a setup with three chains with different values of $q$ or $N$, where one is responsible for the DM stability and interactions, whereas the remaining two give rise to the light-neutrino masses and mixing. 

Notice that the $N \sim R_0$ eigenstates of the two neutrino mass chains could be used to generate the observed baryon asymmetry of the Universe via resonant leptogenesis~\cite{Pilaftsis:1997jf}. If one assumes that the Majorana breaking is universal for two neutrino mass chains, the mass splitting between the two unstable $N \sim R_0$, necessary for resonant leptogenesis, can be generated by the interaction with the clockwork chains, since the exact mass eigenvalues differ slightly from $m_N$ and are function of the $q$ and $N$ of both chains, which are not necessarily the same. By literally repeating the standard argument for leptogenesis, one can figure out that the washout induced by the Yukawa couplings with the SM neutrinos is not large, and therefore leptogenesis is in principle possible. 
%A quantitative analysis of this setup may be given elsewhere.

From the continuum 5D point of view, in the flat spacetime construction given in the previous section, the clockwork suppression is essentially given by the mass $M$. By assuming 3 chains as above, {\em i.e.}~3 fermions $\psi$ in the bulk and 3 chiral partners ({\em i.e.}~3 neutrino chiral partners $R_0$ on the other end of the chain with respect to the SM fields), one could induce 2 non-negligible neutrino masses and account for DM. One of the $R_0$ is essentially the DM particle. The 2 others are necessary to induce 2 non-negligible neutrino masses.
Such a 2-brane setup requires 2 of the bulk fermions to have similar masses (involved in the neutrino mass generation), whereas the third one (related to DM) with a larger mass.  Instead, 
%again if the SM is localized only on the last site brane, 
in the curved spacetime construction of~\cite{Giudice:2016yja}, since the clockwork suppression is universally determined by the metric, such a setup appears not to be feasible (unless one introduces an extra mass term for the bulk field as above).
If one allows the chiral partners to be not all located on the first brane, but placed on different sites (for instance one on the first site for DM and 2 much closer to the other end of the chain for neutrino masses)
%, with $i\neq N$ not too different from $N$, 
more possibilities show up. In this case, in both the flat and curved cases, 2 bulk fermions coupling to different lepton flavour combinations might be enough to generate 2 neutrino masses and account for DM, provided that there are still 3 chiral partners.
%Or alternatively one neutrino mass could be induced from the same bulk fermion as the one associated to DM, if at least one of the $R_i$ with $i$ relatively close to $N$ acquires a soft mass. But in any case if the SM couples only to the last site $R_N$ field, at least a second bulk fermion is necessary to induce a second neutrino mass. This gives an idea of the minimum complexity needed to generate neutrino masses.}
%, if the couplings to the SM lepton are localized on the last site brane. 
A more detailed discussion of this may be given elsewhere.

\paragraph*{\bf Option 2.} If one assumes that the SM fields interact in different flavour combinations with several sites along the chain, a single bulk fermion can be enough to generate two or three nonzero neutrino masses and account for DM, provided that again there are 3 chiral partners (with the one responsible for DM placed far from the 2 others as well as from the SM fields, so that the clockwork suppression for the neutrino masses gets effectively reduced compared to the DM one).
This holds for the flat case considered above, as well as for the curved case considered in \cite{Giudice:2016yja}.
%In order for this to happen, the couplings in at least two different sites must involve 2 different lepton flavour combinations.
%, \emph{i.e.}~the relevant Yukawa couplings need to be not aligned in flavour space. 
%In both the flat and curved spacetime constructions discussed, above and in~\cite{Giudice:2016yja} respectively, this requires at least an extra soft Majorana mass for an intermediate $R_i$ along the chain, so that the clockwork suppression for the neutrino masses gets effectively reduced compared to the DM one.

%\footnote{One chain could be the DM one, with an extra soft Majorana mass for an intermediate $R_i$ along the chain. However, another chain is necessary for inducing at least a second neutrino mass, if only the last site interacts with the SM. If the SM fields interact with several sites along the chain, a single chain might be enough.}

\section{Conclusions}
\label{sec:conc}

In this work we have  proposed an implementation of a WIMP dark matter candidate within the clockwork framework. The candidate is a Majorana singlet particle that
is stable over cosmological time because its
%, unlike most DM scenarios, is not protected from decaying by the imposition of an ad-hoc discrete symmetry. Its 
effective Yukawa couplings to the SM Higgs and leptons are suppressed through a clockwork mechanism. For appropriate choices of the parameters, it can be made very long lived, {\em e.g.} $\tau \gtrsim 10^{26}$ s. The very same particles that suppress this coupling (the clockwork gears) 
%lie at at rather low scale, say in the TeV range, and 
have unsuppressed couplings to the DM candidate, so that its relic abundance may be set by the standard freeze-out mechanism. 

Concretely, as proof of existence, we have considered a simple clockwork chain of fermions, with a single chiral mode at  
%(essentially the DM candidate) at 
one end of the chain, and the SM  leptons at the other end. In Sec.~\ref{sec:pheno} we have studied the basic DM phenomenology ({\em i.e.}~relic abundance, direct detection constraints, collider and low-energy limits and prospects) for various scenarios. These constraints, in particular the relic density one, require the clockwork gears to lie at most at the $\sim$~TeV scale, 
{\em i.e.}~accessible at colliders. Future LFV experiments will also be able to probe efficiently their effects.
Depending on the assumption one makes 
about the nature and relative mass hierarchy of the scalar-sector, many viable candidates may be found. 
%regarding the coupling of the Higgs scalar to the fields along the chain.
 Interestingly, a similar (albeit parametrically distinct) clockwork construction may be relevant for SM neutrino phenomenology and in particular it can explain the smallness of the SM neutrinos, see Sec.~\ref{sec:neutrino}. 
 We pointed out that, in order to have several non-vanishing neutrino masses, one needs several clockwork chains, if only the last site of the chains couples to the SM fields.

Intuitively, the clockwork mechanism imposes that the degrees of freedom at the two edges of the chain have little overlap, a picture that is supported by an embedding of the clockwork chain in a 5D spacetime.  Interestingly, our implementation  departs from the 5D continuum construction of~\cite{Giudice:2016yja}, where the bulk field is massless and the mass gap  between the massless mode and the clockwork gears is generated by a curved metric. In the construction of Sec.~\ref{sec:deconstr}, instead, the metric is flat and the bulk field is massive. The gap between the massless mode, introduced on one of the edges (``branes'')  and the clockwork gears is, in our case, generated by the mass term of the bulk field. Although we have explicitly shown how the clockwork chain can originate from a flat spacetime setup in the fermionic case, a very similar analysis can be carried out for a scalar clockwork: again, one needs a massive bulk scalar and extra brane terms.

%an extra scalar on one brane, the latter parametrically lighter than the bulk one. As long as there is a mass gap between the scalar on one brane and the clockwork gears, the interactions of the light scalar with the SM on the other brane are clockwork-suppressed. 

%%%%%%%%%%%%%%%%%%%%%%%%%%%%%%%%%%%%%%%%%%%%%%%%%%%%%%%%%%%%%%%%%%%%%%

%\vspace{-3mm}

\bigskip\bigskip
\acknowledgements

%\vspace{-2mm}

This work is supported by the FNRS-FRS, the FRIA, the IISN, a ULB-ARC, the Belgian Science Policy (IAP VI-11) and a ULB Postdoctoral Fellowship.


\begin{thebibliography}{99}

%\cite{Hambye:2010zb}
\bibitem{Hambye:2010zb}
  T.~Hambye,
  %``On the stability of particle dark matter,''
  PoS IDM {\bf 2010} (2011) 098
  [arXiv:1012.4587 [hep-ph]].
  %%CITATION = ARXIV:1012.4587;%%
  %31 citations counted in INSPIRE as of 03 Dec 2016
  
 %\cite{'tHooft:1979bh}
%\bibitem{'tHooft:1979bh}
%  G.~'t Hooft,
%  %``Naturalness, chiral symmetry, and spontaneous chiral symmetry breaking,''
%  NATO Sci.\ Ser.\ B {\bf 59} (1980) 135.
%%  doi:10.1007/978-1-4684-7571-5_9
%  %%CITATION = doi:10.1007/978-1-4684-7571-5_9;%%
%  %273 citations counted in INSPIRE as of 03 Dec 2016

 
%\cite{Choi:2015fiu}
\bibitem{Choi:2015fiu}
  K.~Choi and S.~H.~Im,
  %``Realizing the relaxion from multiple axions and its UV completion with high scale supersymmetry,''
  JHEP {\bf 1601} (2016) 149
  [arXiv:1511.00132 [hep-ph]].
  %%CITATION = doi:10.1007/JHEP01(2016)149;%%
  %28 citations counted in INSPIRE as of 03 Dec 2016
   
 %%\cite{Kaplan:2015fuy}
\bibitem{Kaplan:2015fuy}
D.~E.~Kaplan and R.~Rattazzi,
  %``Large field excursions and approximate discrete symmetries from a clockwork axion,''
Phys.\ Rev.\ D {\bf 93} (2016) no.8,  085007
%  doi:10.1103/PhysRevD.93.085007
[arXiv:1511.01827 [hep-ph]].
  %%CITATION = doi:10.1103/PhysRevD.93.085007;%%
  %30 citations counted in INSPIRE as of 03 Dec 2016  
   
\bibitem{Giudice:2016yja}
  G.~F.~Giudice and M.~McCullough,
  %``A Clockwork Theory,''
  JHEP {\bf 1702} (2017) 036
  [arXiv:1610.07962 [hep-ph]].
  
\bibitem{Kehagias:2016kzt}
  A.~Kehagias and A.~Riotto,
  %``Clockwork Inflation,''
  Phys.\ Lett.\ B {\bf 767} (2017) 73
  [arXiv:1611.03316 [hep-ph]].
  
\bibitem{Farina:2016tgd}
  M.~Farina, D.~Pappadopulo, F.~Rompineve and A.~Tesi,
  %``The photo-philic QCD axion,''
  JHEP {\bf 1701} (2017) 095
  [arXiv:1611.09855 [hep-ph]].
  
\bibitem{Ahmed:2016viu}
  A.~Ahmed and B.~M.~Dillon,
  %``Clockwork Composite Higgses,''
  arXiv:1612.04011 [hep-ph].  
  
  %\cite{ArkaniHamed:2001ca}
\bibitem{ArkaniHamed:2001ca}
  N.~Arkani-Hamed, A.~G.~Cohen and H.~Georgi,
  %``(De)constructing dimensions,''
  Phys.\ Rev.\ Lett.\  {\bf 86} (2001) 4757
  [hep-th/0104005].
  %%CITATION = doi:10.1103/PhysRevLett.86.4757;%%
  %652 citations counted in INSPIRE as of 03 Dec 2016

\bibitem{Cline:2013gha}
  J.~M.~Cline, K.~Kainulainen, P.~Scott and C.~Weniger,
  %``Update on scalar singlet dark matter,''
  Phys.\ Rev.\ D {\bf 88} (2013) 055025
   Erratum: [Phys.\ Rev.\ D {\bf 92} (2015) no.3,  039906]
  [arXiv:1306.4710 [hep-ph]].


\bibitem{Akerib:2016vxi}
  D.~S.~Akerib {\it et al.} [LUX Collaboration],
  %``Results from a search for dark matter in the complete LUX exposure,''
  Phys.\ Rev.\ Lett.\  {\bf 118} (2017) no.2,  021303
  [arXiv:1608.07648 [astro-ph.CO]].

  
\bibitem{Aprile:2015uzo}
  E.~Aprile {\it et al.} [XENON Collaboration],
  %``Physics reach of the XENON1T dark matter experiment,''
  JCAP {\bf 1604} (2016) no.04,  027
  [arXiv:1512.07501 [physics.ins-det]].

  



\bibitem{Falkowski:2015iwa}
  A.~Falkowski, C.~Gross and O.~Lebedev,
  %``A second Higgs from the Higgs portal,''
  JHEP {\bf 1505} (2015) 057
  [arXiv:1502.01361 [hep-ph]].

\bibitem{Robens:2015gla}
  T.~Robens and T.~Stefaniak,
  %``Status of the Higgs Singlet Extension of the Standard Model after LHC Run 1,''
  Eur.\ Phys.\ J.\ C {\bf 75} (2015) 104
  [arXiv:1501.02234 [hep-ph]].
  
\bibitem{Chalons:2016jeu}
  G.~Chalons, D.~Lopez-Val, T.~Robens and T.~Stefaniak,
  %``The Higgs singlet extension at LHC Run 2,''
  PoS ICHEP {\bf 2016} (2016) 1180
  [arXiv:1611.03007 [hep-ph]].
  

\bibitem{Beniwal:2015sdl}
  A.~Beniwal, F.~Rajec, C.~Savage, P.~Scott, C.~Weniger, M.~White and A.~G.~Williams,
  %``Combined analysis of effective Higgs portal dark matter models,''
  Phys.\ Rev.\ D {\bf 93} (2016) no.11,  115016
  [arXiv:1512.06458 [hep-ph]].  
  
  %\cite{Pospelov:2007mp}
%\bibitem{Pospelov:2007mp}
%  M.~Pospelov, A.~Ritz and M.~B.~Voloshin,
%  %``Secluded WIMP Dark Matter,''
%  Phys.\ Lett.\ B {\bf 662} (2008) 53
%  [arXiv:0711.4866 [hep-ph]].
%  %%CITATION = doi:10.1016/j.physletb.2008.02.052;%%
%  %410 citations counted in INSPIRE as of 11 Dec 2016


%\bibitem{Lopez-Val:2014jva}
%  D.~L\'opez-Val and T.~Robens,
%  %``Δr and the W-boson mass in the singlet extension of the standard model,''
%  Phys.\ Rev.\ D {\bf 90} (2014) 114018
%  [arXiv:1406.1043 [hep-ph]].


\bibitem{Das:2014jxa}
  A.~Das, P.~S.~Bhupal Dev and N.~Okada,
  %``Direct bounds on electroweak scale pseudo-Dirac neutrinos from $\sqrt s=8$ TeV LHC data,''
  Phys.\ Lett.\ B {\bf 735} (2014) 364
  [arXiv:1405.0177 [hep-ph]].
  
\bibitem{TheMEG:2016wtm}
  A.~M.~Baldini {\it et al.} [MEG Collaboration],
  %``Search for the lepton flavour violating decay $\mu ^+ \rightarrow \mathrm {e}^+ \gamma $ with the full dataset of the MEG experiment,''
  Eur.\ Phys.\ J.\ C {\bf 76} (2016) no.8,  434
  [arXiv:1605.05081 [hep-ex]].
  
\bibitem{mu2e}
 R.~J.~Abrams {\it et al.}  [Mu2e Collaboration],
  %``Mu2e Conceptual Design Report,''
  arXiv:1211.7019 [physics.ins-det]. 

\bibitem{comet} 
Y.~G.~Cui {\it et al.}  [COMET Collaboration],
  %``Conceptual design report for experimental search for
  % lepton flavor violating mu- - e- conversion at sensitivity
  % of 10**(-16) with a slow-extracted bunched proton beam (COMET),''
  KEK-2009-10. 

\bibitem{Deppisch:2015qwa}
  F.~F.~Deppisch, P.~S.~Bhupal Dev and A.~Pilaftsis,
  %``Neutrinos and Collider Physics,''
  New J.\ Phys.\  {\bf 17} (2015) no.7,  075019
  [arXiv:1502.06541 [hep-ph]].
  
  %\cite{Aisati:2015vma}
\bibitem{Aisati:2015vma}
  C.~El Aisati, M.~Gustafsson and T.~Hambye,
  %``New Search for Monochromatic Neutrinos from Dark Matter Decay,''
  Phys.\ Rev.\ D {\bf 92} (2015) no.12,  123515
  [arXiv:1506.02657 [hep-ph]].
  %%CITATION = doi:10.1103/PhysRevD.92.123515;%%
  %15 citations counted in INSPIRE as of 17 Dec 2016

  %\cite{Creutz:1984mg}
\bibitem{Creutz:1984mg}
  M.~Creutz,
  ``Quarks, gluons and lattices,''
 Cambridge Univ. Press (1985).
  %%CITATION = INSPIRE-205398;%%
  %17 citations counted in INSPIRE as of 09 Dec 2016

%\cite{Kaplan:2009yg}
\bibitem{Kaplan:2009yg}
  D.~B.~Kaplan,
  %``Chiral Symmetry and Lattice Fermions,''
  arXiv:0912.2560 [hep-lat].
  %%CITATION = ARXIV:0912.2560;%%
  %27 citations counted in INSPIRE as of 11 Dec 2016

\bibitem{Pilaftsis:1997jf}
  A.~Pilaftsis,
  %``CP violation and baryogenesis due to heavy Majorana neutrinos,''
  Phys.\ Rev.\ D {\bf 56} (1997) 5431
  [hep-ph/9707235].

\end{thebibliography}
\end{document}